**Quantum Computing Using Single Photons and the Zeno Effect**

J.D. Franson, B.C. Jacobs, and T.B. Pittman

Johns Hopkins University

Applied Physics Laboratory

Laurel, MD 20723

*Abstract:*

We show that the quantum Zeno effect can be used to suppress the failure events that would otherwise occur in a linear optics approach to quantum computing. From a practical viewpoint, that would allow the implementation of deterministic logic gates without the need for ancilla photons or high-efficiency detectors. We also show that the photons can behave as if they were fermions instead of bosons in the presence of a strong Zeno effect, which leads to a new paradigm for quantum computation.

## I.        Introduction

A linear optics approach to quantum computing [1] would have a number of practical advantages.  Several devices of that kind have been experimentally demonstrated [2-7], including controlled-NOT (CNOT) logic gates and a small-scale quantum circuit for photonic qubits [8].  These devices are probabilistic in nature, however, and large numbers of ancilla photons would have to be generated in entangled states and detected with high efficiency in order to minimize the inherent failure rate.  Here we show that the quantum Zeno effect [9-15] can be used to suppress those failure events, which would eliminate the need for ancilla photons and high-efficiency detectors.  We also show that the photons can behave like non-interacting fermions instead of bosons in the presence of a strong Zeno effect, which is of fundamental interest.

All of the failure events in our original linear optics CNOT gate [2, 3] correspond to the emission of more than one photon into the same optical fiber.  In the quantum Zeno effect, a randomly-occurring event can be suppressed by frequent measurements to determine whether or not the event has occurred.  The basic idea of our approach is to use the Zeno effect to suppress the emission of more than one photon into an optical fiber, which would eliminate the source of failures.  Although these techniques can be applied directly to our original CNOT gate, it is simpler to implement a $\sqrt{SWAP}$ [16, 17] gate instead.  The motivation for the approach is discussed in Section II, along with a proposed implementation of a $\sqrt{SWAP}$ gate using two coupled optical fibers and the Zeno effect.

The performance of the proposed $\sqrt{SWAP}$ gate is analyzed in Section III for the idealized case in which a series of measurements are made in order to determine the presence of two or more photons in the same optical fiber.  The system is assumed to propagate in accordance with Schrodinger's equation between the measurements and to be reduced (collapse) after each measurement as required by quantum measurement theory.  The results of these calculations show that the coupled optical fiber device does function as a $\sqrt{SWAP}$ gate in the limit of a large number of such measurements, aside from a phase factor that has no effect on our ability to perform universal quantum computation.

As is usually the case in the quantum Zeno effect, no actual observations or measurements are required.  Instead, equivalent results can be obtained using strong two-photon absorption within the optical fibers.  Section IV describes a density matrix calculation that was used to analyze this approach, which gives very nearly the same results as the discrete measurements of Section III.  A number of practical considerations are also considered here, including an estimate of the achievable rate of two-photon absorption in optical fibers and ways to minimize the effects of single-photon scattering and absorption.

The presence of a Zeno effect of this kind inhibits the emission of more than one photon into the same optical mode.  As a result, the photons are forced to obey the Pauli exclusion principle as if they were fermions instead of bosons. This situation is discussed in more detail in Section V, where it is shown that the evolution of the state vector for photons in the presence of a strong Zeno effect is exactly the same as it would be for non-interacting fermions (without the Zeno effect). It is also shown that the quantum interference properties of the photons are those of fermions instead of bosons, and that the time-averaged field operators obey anti-commutation relations.

The ability to perform universal quantum computation using non-interacting fermions would appear to contradict the well-known no-go theorems [18, 19] for non-interacting fermions. This situation is discussed in Section VI, where it is shown that the ability to perform quantum logic operations in this way depends on the fact that the photons can be forced to behave like non-interacting fermions in one part of a circuit and like non-interacting bosons in other parts of the circuit. This can be viewed as a new paradigm for quantum computation.

A scalable approach to quantum computing would require the errors in the quantum logic gates to be below the threshold for quantum error correction. Section VII discusses a two-qubit encoding [1] that has an error threshold of 1/4 when used in conjunction with our proposed Zeno gates. The use of such an encoding would allow scalable performance to be achieved even when the two-photon absorption rate is limited.

We conclude with a summary in section VIII, including a discussion of the prospects for using Zeno gates for practical applications.

## II.    Motivation and $\sqrt{SWAP}$ gate

The implementation of quantum logic gates has always been one of the main challenges in an optical approach to quantum computing. Although logic operations are inherently nonlinear, Knill, Laflamme, and Milburn (KLM) showed that they could be performed using linear optical elements, additional photons (ancilla), and post-selection and feed-forward control based on the results of measurements on the ancilla photons [1]. Roughly speaking, the devices are designed in such a way that the quantum-mechanical measurement process projects out a final state corresponding to the desired logical output. Several groups have now demonstrated a number of logic gates of this kind [2-8].

Despite the rapid progress in the development of probabilistic quantum logic gates using linear optical elements, their reliance on ancilla photons poses a serious challenge. The probability of a failure event or error scales as $1/n$ in the original KLM approach, where $n$ is the number of ancilla photons, while it scales as $1/n^2$ in an alternative approach that we have suggested [20]. The ancilla can be generated in the necessary entangled initial state using elementary logic gates (post-selection), but the efficiency of an approach of that kind decreases exponentially with increasing values of $n$ [21]. In addition to generating the ancilla photons, they must be detected with high efficiency in order to avoid errors in the output of the logic gate. With this in mind, we have recently been considering the possibility of hybrid approaches in which the need for large numbers of ancilla photons could be reduced or eliminated by combining linear optics techniques with some amount of nonlinearity. The Zeno gates described here are an example of a hybrid approach of that kind.

The origin of the failure events in our original linear optics CNOT gates [22] can be understood by considering the implementation shown in Fig. 1. This is a relatively simple device consisting of two polarizing beam splitters, two polarization-sensitive detectors, and a pair of entangled ancilla photons used as a resource. It can be shown that this device will produce the desired CNOT operation (aside from any technical errors) if one and only one photon is detected in each of the detectors, which occurs with a probability of $1/4$. The output of the device must be rejected, however, if two photons exit in any of the four output modes. Our most recent experiments have all been implemented in optical fibers, so that the failure events correspond to the emission of two photons into the same optical fiber at one or more of the beam splitters (fiber couplers). The basic idea is to use the quantum Zeno effect to suppress the emission of more than one photon into the same fiber core, which would prevent the failure events and produce a deterministic logic gate that succeeds 100% of the time.

In order for the Zeno effect to suppress the build-up of undesired probability amplitudes, it is necessary to apply a series of measurements on a time scale that is small compared to the time required for the emission of two photons into the same optical fiber. This requires that the beam splitter operation be performed continuously over a relatively large time scale, which can be accomplished using the dual-core optical fiber geometry shown in Figure 2. Here the cores of two optical fibers are assumed to run parallel to each other in close proximity over some distance L. The photons are assumed to occupy only the fundamental transverse mode of the fiber core with a fixed linear polarization. The overlap of the evanescent fields of the two cores will gradually couple a photon from one optical fiber core into the other in a manner that is analogous to tunneling. Similar devices are available commercially and are used as optical fiber couplers. They are equivalent to free-space beam splitters, aside from the gradual nature of the transition from one core to the other.

Although it is possible to use the Zeno effect to suppress the failure events for the CNOT gate shown in Fig. 1, it is simpler to implement another logic gate that only requires a single beam splitter. The resulting device is similar to the square-root of SWAP gate illustrated in Fig. 3a. The SWAP operation interchanges the values of two input qubits, while the $\sqrt{SWAP}$ gate is defined as producing the SWAP operation when applied twice or squared, as shown in the figure. It is well-known that the $\sqrt{SWAP}$ operation is universal for quantum computation when combined with single-qubit operations [16, 17].

Figure 3b suggests that it may be possible to implement a $\sqrt{SWAP}$ gate directly using the coupled-fiber geometry of Fig. 2. If the length L of the interaction region is chosen properly, a photon incident in one fiber core will be transferred to the other core with 100% probability. This will implement the $SWAP$ operation if the absence of a photon is assumed to represent a logical value of 0 while the presence of a photon represents a logical value of 1. If we reduce the length of such a device by a factor of two to $L_{1/2} = L/2$, as illustrated by the dashed line in Fig. 3b, then we will produce a device that gives the $SWAP$ operation when applied twice, or squared. This corresponds to the definition of a $\sqrt{SWAP}$ gate, which suggests that such an operation could be produced using a length $L_{1/2}$ of coupled fibers.

Such a situation is too good to be true, as one might expect. We will show in the next section that the coupled-fiber device of Figure 3b does implement a $\sqrt{SWAP}$ operation correctly if a total of 0 or 1 photons are input to the device. Incorrect results are obtained, however, if a photon is present in both of the input modes, which corresponds to a logical input of 1 for both qubits. In that case, there is some probability that both photons will exit the device in the same fiber core, which corresponds to an error state since only 0 or 1 photons represents a valid logical output. In fact, quantum interference effects ensure that both photons will always exit such a device in the same fiber core, which is equivalent to the well-known Hong-Ou-Mandel dip [23] in coincidence measurements using free-space beam splitters. Errors of that kind are analogous to the failure events in our original CNOT gate, and we will see in the next section that the Zeno effect can be used to suppress them.

### III. Discrete measurements

In this section, the coupling of photons from one fiber core into the other will be analyzed in more detail. For a single incident photon, the behavior of the coupled system will be found to be analogous to the Rabi oscillations of a two-level atom. The ability of the Zeno effect to suppress errors in which two photons are emitted into the same fiber core will then be investigated by assuming that a series of discrete measurements are made to determine whether or not two photons are present in the same core. A more realistic implementation involving two-photon absorption will be analyzed in Section IV.

In the limit of weak coupling, the Hamiltonian for the system described above can be written in the form

$$\hat{H} = \sum_k [\hbar \omega_k (\hat{a}_{k1}^{\dagger} \hat{a}_{k1} + \hat{a}_{k2}^{\dagger} \hat{a}_{k2}) + \varepsilon (\hat{a}_{k1}^{\dagger} \hat{a}_{k2} + \hat{a}_{k2}^{\dagger} \hat{a}_{k1})] \tag{1}$$

Here $k$ is the longitudinal k-vector describing the propagation of the photons down the fiber, $\hbar$ is Planck's constant divided by $2\pi$, $\omega_k$ is the angular frequency of the photons, the operators $\hat{a}_{k1}^{\dagger}$ and $\hat{a}_{k2}^{\dagger}$ create photons in each of the fibers, and the parameter $\varepsilon$ determines the strength of the coupling between the two fibers. The zero-point energy has no effect in this system and was omitted from Eq. (1).

If the bandwidth is sufficiently small that $\varepsilon$ is approximately independent of the value of $k$, then we can work in the interaction picture by writing $\hat{H} = \hat{H}_0 + \hat{H}'$, where $\hat{H}_0$ includes the energies of the photons. The unperturbed states then correspond to photon wave packets propagating freely down one of the fibers, while the perturbation Hamiltonian has the form

$$\hat{H}' = \varepsilon(\hat{a}_1^\dagger \hat{a}_2 + \hat{a}_2^\dagger \hat{a}_1) \tag{2}$$

Here the operators $\hat{a}_1^\dagger$ and $\hat{a}_2^\dagger$ create photons in the corresponding wave-packet states in one of the two fiber cores.

An arbitrary input state can be expressed as a superposition of basis states in the computational basis. We will calculate the time evolution of each of the basis states individually. From the linearity of quantum mechanics, those results can be used to obtain all of the elements of the unitary transformation matrix that describes the operation of the device. This procedure is equivalent to conventional S-matrix theory.

We first consider the case of a single incident photon, where the Hamiltonian of Eq. (2) is equivalent to that of a two-level atom coupled to a classical electromagnetic field. Here the most general state of the system is a superposition of the two basis states $|\psi_1\rangle = \hat{a}_1^\dagger |0\rangle$ and $|\psi_2\rangle = \hat{a}_2^\dagger |0\rangle$, where $|0\rangle$ is the vacuum state. (This forms a subspace of the original Hilbert space.) In that basis, the perturbation Hamiltonian is given by

$$\hat{H}' = \begin{pmatrix} 0 & \varepsilon \\ \varepsilon & 0 \end{pmatrix} \tag{3}$$

For the case in which a single photon is incident in path 1, Schrodinger's equation has the solution

$$|\psi(t)\rangle = \begin{pmatrix} Cos(t) \\ -iSin(t) \end{pmatrix} \tag{4}$$

For simplicity, the time $t$ has been expressed in units given by $\hbar / \varepsilon$. This solution is equivalent to a Rabi oscillation in which a photon wave packet is coupled back and forth between the two fiber cores in a periodic manner as it propagates down the system, as illustrated in Fig. 4. It can be seen that the system couples completely into the opposite path after a time $t = \pi / 2$ and a corresponding distance $L = ct = \pi c \hbar / 2\varepsilon$. At half that distance, $L_{1/2} = L / 2$, the solutions to Schrodinger's equation correspond to that of a $\sqrt{SWAP}$ operation for a single incident photon without any need for the Zeno effect. A similar solution exists for the case in which a single photon is incident in path 2.

The situation is more complicated, however, for the case in which a single photon is incident in each of the optical fiber cores, which corresponds to a logical value of 1 for both qubits. In addition to the initial state of $|\psi_{11}\rangle = \hat{a}_1^\dagger \hat{a}_2^\dagger |0\rangle$, there is also a coupling into the two-photon states $|\psi_{20}\rangle = \hat{a}_1^{\dagger 2}|0\rangle/\sqrt{2}$ and $|\psi_{02}\rangle = \hat{a}_2^{\dagger 2}|0\rangle/\sqrt{2}$. In that basis, the perturbation Hamiltonian has the form

$$\hat{H}' = \sqrt{2}\varepsilon \begin{pmatrix} 0 & 1 & 1 \\ 1 & 0 & 0 \\ 1 & 0 & 0 \end{pmatrix} \qquad (5)$$

Now Schrodinger's equation has the solution

$$|\psi(t)\rangle = \begin{pmatrix} Cos(2t) \\ -iSin(2t)/\sqrt{2} \\ -iSin(2t)/\sqrt{2} \end{pmatrix} \qquad (6)$$

The probability $P_{11}(t)$ that the system is in the initial state $|\psi_{11}\rangle$ with one photon in each path at time $t$ is simply

$$P_{11}(t) = Cos^2(2t) \qquad (7)$$

It can be seen that $P_{11} = 0$ after the photons have traveled a distance of $L_{1/2}$ ($t = \pi/4$), which implies that both photons are located in the same path at that point. Since a coupled fiber device of length $L_{1/2}$ is equivalent to a free-space 50/50 beam splitter, this result is consistent with the well-known Hong-Ou-Mandel interference effect [23] as will be discussed in more detail in Section V. This situation corresponds to an error state at the output of the device, since the only logical states correspond to either 0 or 1 photon in each of the fiber cores.

In order to investigate the ability of the Zeno effect to suppress the emission of two photons into the same fiber core, we assumed that a total of $N$ discrete measurements were made during the time $\Delta t = L_{1/2}/c$ that the photons spend in the coupled-fiber region. Each measurement was assumed to be able to identify the presence of two photons in the same fiber core, in which case the qubits were destroyed and the operation of the device was considered to be a failure. The absence of two photons in the same fiber core was assumed to project the state vector into the orthogonal subspace (consisting of state $|\psi_{11}\rangle$), as required by the measurement postulate of quantum mechanics. Experimental techniques for implementing measurements of that kind will be discussed in Section IV.

The effects of such a sequence of measurements were calculated by propagating the initial state $|\psi_{11}\rangle$ up to the time of the first measurement at $t = \Delta t/N = \pi/4N$ using

Schrodinger's equation. From Eq. (7), the probability $P_F$ of a failure event in which two photons are detected in the same fiber core is given by

$$P_F = 1 - P_{11}(\Delta t / N) = 1 - Cos^2(2\Delta t / N) = 1 - Cos^2(\pi / 2N) \tag{8}$$

The other possibility is that the system will be successfully projected back into the initial state, which occurs with a probability of $P_S = Cos^2(\pi / 2N)$. After this process has been repeated $N$ times, the overall probability of a successful outcome (no failure events) is $P_S = \cos^{2N}(\pi / 2N)$.

The failure probability $P_E$ is plotted as a function of the number of measurements $N$ in Fig. 5. It can be seen that $P_E = 1$ if a single measurement is made at the end of the process, as is consistent with the Hong-Ou-Mandel effect. But the error probability approaches zero in the limit of large $N$, where $P_E = \pi^2 / 4N$, as would be expected from the quantum Zeno effect [9-15]. The fact that the error is only inversely proportional to $N$ may seem to imply that large values of $N$ would be required in order to meet the threshold for quantum error correction. But a simple two-qubit encoding [1] can be used to achieve small error rates even for moderate values of $N$, as is described in Section VII.

By definition [16, 17], a $\sqrt{SWAP}$ gate applies a unitary transformation in the computational basis given by

$$\sqrt{SWAP} = \begin{bmatrix} 1 & 0 & 0 & 0 \\ 0 & (1+i)/2 & (1-i)/2 & 0 \\ 0 & (1-i)/2 & (1+i)/2 & 0 \\ 0 & 0 & 0 & 1 \end{bmatrix} \tag{9}$$

In order to facilitate comparison with a standard $\sqrt{SWAP}$ operation, we assumed that a phase shift of $\pi / 4$ was inserted in each of the output ports of the coupled-fiber device shown in Fig. 3b. In the limit of large $N$, the final state of the system corresponds to a unitary transformation given by

$$\sqrt{SWAP}' = \begin{bmatrix} 1 & 0 & 0 & 0 \\ 0 & (1+i)/2 & (1-i)/2 & 0 \\ 0 & (1-i)/2 & (1+i)/2 & 0 \\ 0 & 0 & 0 & i \end{bmatrix} \tag{10}$$

Aside from the calculations described above, these results are apparent for the case of a single incident photon, where they correspond to the usual results for Rabi oscillations. They are also apparent for the case of two incident photons, since the Zeno effect essentially eliminates the coupling between the fiber cores while the fixed phase shift of $\pi/4$ in each path is responsible for the factor of $i$ in the lower diagonal.

It can be seen that the unitary transformation of Eq. (10) differs from that of a standard $\sqrt{SWAP}$ operation by a factor of $i$ in the lowest diagonal element. As a result, we will refer to this operation as $\sqrt{SWAP'}$, and we will denote its square by

$$SWAP' \equiv (\sqrt{SWAP'})^2 = \begin{bmatrix} 1 & 0 & 0 & 0 \\ 0 & 0 & 1 & 0 \\ 0 & 1 & 0 & 0 \\ 0 & 0 & 0 & -1 \end{bmatrix} \tag{11}$$

The $SWAP'$ operation of Eq. (11) differs from a standard $SWAP$ by a factor of $-1$ in the lowest diagonal element.

We conclude this section by describing how the SWAP' operation from the coupled fiber device of Fig. 3b can be used to implement a CNOT logic gate. The SWAP' of Eq. (11) differs from a controlled-Z operation (controlled phase gate) only by the fact that two of its terms are off the diagonal. Those terms can be put back on the diagonal by following the SWAP' operation by a standard $SWAP$, and the combined effect of those two operations is to implement a controlled-Z gate as illustrated in Fig. 6b. For photonic qubits, the $SWAP$ can be applied by simply interchanging two optical fibers, as shown in Fig. 6a. The contolled-Z operation of Fig. 6 is universal for quantum computation when combined with single-qubit operations. In particular, it can be combined with two Hadamards (beam splitter operations) to implement a CNOT gate. Thus the $SWAP'$ operation allows any quantum computation to be performed when combined with single-photon operations and the interchange of optical fibers.

## IV.    Two-photon absorption

In the previous section, it was shown that a sequence of $N$ measurements could be used to suppress failure events in which two photons are emitted into the same optical fiber core. As is usually the case in the Zeno effect, no actual measurements or observations are required. Instead, the system of interest can be coupled to a second system in such a way that subsequent measurements on the second system could provide the same information. In the situation of interest here, that could be accomplished by inserting one or more atoms into the cores of the optical fibers in such a way that the atoms can absorb two photons but not just one. Subsequent measurements could determine whether or not the atoms were left in an excited state, which would indicate the presence of more than one photon in the same core. As a result, one would expect that strong two-photon absorption should give the same error suppression as the sequence of discrete measurements considered in the previous section.

In order to investigate this possibility, we assumed that the two-photon states $|\psi_{20}\rangle$ and $|\psi_{02}\rangle$ were absorbed at a rate of $1/\tau_D$, where $\tau_D$ is the corresponding decay time. It was further assumed that the single-photon states $|\psi_1\rangle$ and $|\psi_2\rangle$ were unaffected by this process, so that we only needed to consider the case of two incident photons with the interaction Hamiltonian given by Eq. (5). In addition to these three states, it was assumed that there was a quasi-continuum of excited atomic states into which the two-photon states could decay. The process could then be described by a density-matrix calculation in which the rate of change of the density matrix elements due to the interaction Hamiltonian $\hat{H}'$ was given as usual by

$$\dot{\rho}_{ij} = \frac{1}{i\hbar}\sum_k (H_{ik}\rho_{kj} - \rho_{ik}H_{kj}) \tag{12}$$

In addition to the Hamiltonian evolution of Eq. (12), the diagonal density-matrix elements $\rho_{dd}$ associated with the two-photon states were assumed to have an additional rate of change due to two-photon absorption given by

$$\dot{\rho}_{dd} = -\frac{1}{\tau_D}\rho_{dd} \tag{13}$$

This coupled set of differential equations was solved using Mathematica.

The solid line in Fig. 5 shows the results of a density matrix calculation of this kind. In order to allow both sets of calculations to be plotted on the same scale, the parameter $N$ was defined in this case as $N = \Delta t / 4\tau_D$. It can be seen that strong two-photon absorption can inhibit the emission of two photons into the same mode in very nearly the same way as a series of discrete measurements. In the limit of small $\tau_D$, this density matrix calculation gives the same unitary transformation as the discrete measurements of Section III, namely the $\sqrt{SWAP}'$ of Eq. (10).

It should be emphasized that strong two-photon absorption does not imply that there will be a large rate of decoherence due to rapid absorption of photon pairs. On the contrary, the existence of a strong two-photon absorption mechanism will prevent the emission of pairs of photons into the same mode in the first place, and there need be no dissipation or decoherence associated with this process in the limit of small $\tau_D$.

The Zeno effect has previously been investigated as a way to suppress Rabi oscillations in two-level systems [10-12, 15], and the same factor of 4 (in $N = \Delta t / 4\tau_D$) has previously been reported [13, 14]. The use of linear dissipation in cavity QED systems has also been proposed by Beige et al. [25-27] as a method of suppressing unwanted error mechanisms. Although there is an obvious connection between these earlier approaches and our Zeno gate proposal, we are not suppressing the usual Rabi oscillations of Fig. 4. Instead, nonlinear dissipation (two-photon absorption) is used to suppress the coupling into the undesired states with two photons in the same fiber core.

Nonlinear effects such as two-photon absorption are commonly assumed to be small at single-photon intensities. However, the possibility of strong two-photon absorption in optical fibers can be understood by comparison with nonlinear effects at low intensities in cavity QED experiments [28]. For example, a single-photon wave packet from a 100 fs pulsed source would have a length of approximately 30 $\mu m$ and would be confined in an area $\sim 1$ $\mu m$ in diameter, which corresponds to a mode volume that is much smaller than that commonly used in cavity QED experiments. The concentration of the photon energy into such a small volume can produce relatively large electric fields even at the single-photon level and relatively large nonlinearities, including two-photon absorption, can be expected as a result.

In order to estimate the rate of two-photon absorption in optical fibers, we considered the case of hollow fiber cores containing three-level atoms as illustrated in Figs. 7 and 8. The upper atomic level was assumed to decay at a rate of $1/\tau_C$ due to collisions. At the atomic densities at which Zeno gates are expected to operate, $\tau_C \ll \tau_R$, where $\tau_R$ is the radiative lifetime of the upper atomic level due to spontaneous emission. The photon wave packets were assumed to be gaussians with a length (one standard deviation) of $L_P$. The energy of the photons was assumed to be detuned by an amount $\delta$ from the energy of the intermediate atomic state, as illustrated in Fig. 8b, and the detuning was assumed to be sufficiently large that the virtual transition could be described by an effective matrix element in the usual way. For simplicity, the field energy was assumed to be approximately uniform over the area A of the hollow cores. The two-photon absorption rate was then calculated using a density matrix approach similar to that of Eqs. (12) and (13).

With these approximations, the density matrix calculation gave a rate $R_2$ for two-photon absorption that can be written as

$$R_2 = \sqrt{\frac{2}{\pi}} N_A f_\delta f_C f_P \frac{\sigma_0}{A} \frac{1}{\tau_R} \tag{14}$$

Here $\sigma_0$ is the resonant cross-section for the absorption of a single photon. This cross-section can be comparable to the area $A$ of the optical fiber cores, as illustrated in Fig. 7, since it is on the order of the square of the wavelength $\lambda$ [29]:

$$\sigma_0 = \frac{3}{2\pi} \lambda^2 \tag{15}$$

$N_A$ is the number of atoms in a length $L_R = c\tau_R$ of optical fiber and the factors of $f_\delta$, $f_C$, and $f_P$ take into account the effects of detuning, atomic collisions, and the length of the wave packets, respectively, as described below. Two-photon absorption is well understood [30], and Eq. (14) simply casts $R_2$ in a form that is useful for optical fiber applications.

Detuning the photons from the energy of atomic level 2 reduces $R_2$ by a factor of

$$f_\delta = \left( \frac{M_{21}}{\delta} \right)^2 \tag{16}$$

where $M_{21}$ is the atomic matrix element for a transition from the ground state to the first excited state. Although this factor is relatively small, it can be offset by the large value of $N_A$ if we choose the number of atoms to be given by $N_A = 1/f_\delta$. A large value of $N_A$ is also desirable because the atomic density will become nearly uniform, and a perfectly uniform medium does not scatter any non-resonant photons.

The factor $f_C$ represents the fact that collisions will increase the linewidth of the atomic transition and thereby reduce the rate of transitions on resonance. It is given by

$$f_C = \tau_C / \tau_R \tag{17}$$

This factor can range from 0.1 to 0.01 for typical atomic vapor transitions. The factor $f_P$ reflects the fact that nonlinear absorption is proportional to the square of the field intensity, which increases as $L_P$ is decreased and the electromagnetic energy of a photon is concentrated into a smaller volume. This factor is given by

$$f_P = c\tau_R / L_P \tag{18}$$

The increased atomic linewidth due to collisions allows the length of the photon wave packets to be reduced to $L_P \sim c\tau_C$, in which case the product $f_C f_P \sim 1$.

For an optical fiber with a diameter of $0.78 \times \lambda$, the factor of $\sigma_0 / A$ in Eq. (14) is also equal to unity. Although commercially-available optical fibers have diameters that are typically somewhat larger than this, custom-made fibers can be fabricated with core diameters of this size. All of the factors in Eq. (14) then cancel out and the net two-photon absorption rate reduces to

$$R_2 \sim 1/\tau_R \tag{19}$$

This corresponds to a two-photon absorption length $l_2$ ($1/e$ distance) of $l_2 \sim c\tau_R$. It can be seen that choosing $f_\delta = 1/N_A$ in Eq. (16) eliminated the matrix elements and the detuning from the calculations, which is responsible for the simple form of Eqs. (14) and (19).

Aside from the density matrix calculations described above, these results can be qualitatively understood by comparison with the idealized case in which both photons are on resonance, the effects of collisions and Doppler shifts can be neglected, and there is a single atom in each fiber core, as illustrated in Figs. 7 and 8a. If the diameter of the optical fiber core is comparable to $\lambda$, the probability that a resonant photon will be absorbed by a single atom in the core is approximately given by $\sigma_0 / A \sim 1$. Once one photon has been absorbed, the atom will be left in an excited state with a similar cross-section as illustrated in Fig. 8a, so that a second resonant photon could then be absorbed with a probability that is also on the order of unity. In order to satisfy the resonance condition, the length of the wave packets must be comparable to $c\tau_R$. Both photons would then be absorbed with a probability on the order of unity during the time required for the wave packets to pass by an atom, namely $\tau_R$. That gives a two-photon absorption rate of $1/\tau_R$, which is the origin of the last factor in Eq. (14). For the non-ideal case, the effects of detuning are included in the factor of $f_\delta$ which is familiar from perturbation theory, while the expressions for $f_C$ and $f_P$ are equally apparent. The density matrix calculation agrees with this simple argument to within a factor of $\sqrt{2/\pi}$.

Typical values of $l_2$ based on Eq. (19) would be on the order of 5 m in the visible region of the spectrum. If the error probability per gate operation is required to be less than some value $P_E$, then the results of Figure 5 would require that the $\sqrt{SWAP}$ gates have a length $L_{1/2}$ that is greater than $l_2$ by a factor of approximately $1/P_E$. As we mentioned earlier, the two-bit encoding of Section VII allows the use of relatively large values of $P_E$. As a result, the required length of the devices should only be a few times larger than $l_2$.

As a practical matter, the main challenge in implementing devices of this kind will probably be the losses due to single-photon absorption or scattering. Under most conditions, the single-photon scattering might be expected to be larger than the two-photon absorption. For example, a frequent comment is that there will be a "tail" for single-photon scattering that falls off as $1/\delta^2$, and the effects of this tail may be larger than the two-photon absorption even for large detunings. There are several reasons why that need not be the case, however. First of all, the small mode volume greatly increases nonlinear effects such as two-photon absorption as compared to single-photon loss, as we mentioned above. In addition, there is no scattering or absorption in the idealized limit of a perfectly uniform medium with no impurities. In the latter case, the "tail" in the single-photon scattering is suppressed by interference between the scattering amplitudes from different atoms. As an example, quantum key distribution depends on the fact that single photons can travel through kilometers of optical fiber with very little loss, even though they interact strongly enough with the atoms in the optical fiber to produce phase shifts on the order of $10^9$ radians/km.

In addition to the effects of a uniform medium, single-photon scattering can be reduced using electromagnetically-induced transparency (EIT), which has already been suggested as a method for producing relatively large two-photon absorption rates [31]. There may also be other mechanisms for two-photon absorption, such as the four-wave mixing process illustrated in 8c, that can produce higher two-photon scattering rates than the simple process considered above.

Finally, the required length of the coupled fiber device can be greatly reduced if mirrors are used at each end of the fibers to produce resonant cavities. If $f$ is the finesse of the cavities, then the required length of dual-core fiber scales as $1/f^2$ and the single-photon loss scales as $1/f$. For example, a finesse of 100 would reduce the characteristic length of the devices from ~5 m to less than 1 mm. We believe that devices of this kind are feasible and an initial experiment is in progress.

### V.      Fermionic behavior of photons

As we have seen above, a strong Zeno effect can prevent two photons from occupying the same mode, which is analogous to the Pauli exclusion principle for fermions. In this section, we will consider the possibility of fermionic behavior of the photons in more detail. We note that the quantum interference properties of the photons under these conditions are those of fermions instead of bosons. We also show that the dynamic evolution of the system is the same as that of non-interacting fermions, and that the time-averaged creation and annihilation operators obey anti-commutation relations instead of commutators in the limit of a strong Zeno effect.

It is well known that the quantum interference effects responsible for the Hong-Ou-Mandel dip are due to the bosonic nature of photons; the fact that the photons always emerge in the same output port of a beam splitter can be viewed as an extreme example of photon bunching [32, 33]. Electrons or other fermions would give just the opposite result, with both particles always exiting from different output ports [32, 33]. This difference in behavior can be traced to the fact that the exchange of two identical fermions multiplies the state vector by a factor of −1, whereas the exchange of two bosons gives a factor of +1, which converts an interference maximum to a minimum. It can be seen from Figure 5 that the properties of the photons, at least as far as these interference effects are concerned, gradually change from that of a boson to that of a fermion as the strength of the Zeno effect is increased.

An entangled pair of photons in a $\Phi^{(-)}$ Bell state will also emerge into different output ports when incident on a beam splitter [34]. But that is a characteristic property of bosons, not fermions, since the spatial part of the state vector must be anti-symmetric if the angular momentum (polarization) part of the state vector is anti-symmetric. Here we are considering a single polarization (spin) mode, in which case the predicted interference effects are those of a fermion.

The fermionic properties of the photons can be further understood by comparing the results obtained above with what would be expected if true fermions (with no Zeno effect) were incident upon the coupled wave guide device of Section III. Once again, we first consider the case of a single incident particle, where there are only two independent states of the system as in Eq. (3). The Hamiltonian for the system is given by Eq. (2) for either fermions [18] or bosons, and it corresponds to tunneling between the two wave guides in either case. The matrix elements of the Hamiltonian can now be calculated using the anti-commutation relations

$$\{\hat{b}_i, \hat{b}_j^\dagger\} = \hat{b}_i \hat{b}_j^\dagger + s\hat{b}_j^\dagger \hat{b}_i = \delta_{ij} \tag{20}$$

Here the parameter $s = 1$ for fermions, but we have written it as a variable in order to facilitate comparison with the corresponding results for bosons using commutation relations ($s = -1$). The operators $\hat{b}_i^\dagger$ and $\hat{b}_j^\dagger$ create a fermion in the corresponding wave guide.

If $|i\rangle = \hat{b}_i^\dagger |0\rangle$ and $|j\rangle = \hat{b}_j^\dagger |0\rangle$ denote the two basis states containing a single particle, then the off-diagonal matrix elements of the Hamiltonian can be calculated as follows:

$$\langle i | H' | j \rangle = \varepsilon (\hat{b}_i^\dagger |0\rangle)^\dagger \hat{b}_i^\dagger \hat{b}_j (\hat{b}_j^\dagger |0\rangle) = \varepsilon \langle 0 | \hat{b}_i \hat{b}_i^\dagger \hat{b}_j \hat{b}_j^\dagger |0\rangle \tag{21}$$

Using Eq. (20) and the fact that $\hat{b}_i |0\rangle = 0$ and $\hat{b}_j |0\rangle = 0$ reduces the matrix element to

$$\langle i | H' | j \rangle = \varepsilon \langle 0 | \hat{b}_i \hat{b}_i^\dagger |0\rangle = \varepsilon \langle 0 | 0 \rangle = \varepsilon \tag{22}$$

It can be seen that the matrix elements are the same for fermions or bosons (either sign of $s$) for the case of a single particle, as might have been expected. Given the one-to-one correspondence between the single-particle states and the matrix elements, it follows that the solutions to Schrodinger's equation have the same form for the single-particle case whether we are considering fermions or bosons.

Next we consider the case in which two fermions are incident, where the use of commutators vice anti-commutators might be expected to play a role. In the case of fermions, however, only a single state is possible in which one fermion propagates in each wave guide, and $H'$ cannot couple the system to any other state. As a result, the system simply propagates in the initial state with a phase shift that is determined by the energy of the particles. But exactly the same situation holds for the case of two incident photons in the presence of a strong Zeno effect, which effectively eliminates the coupling to states with two photons in the same wave guide. As a result, the system propagates in the initial state with the same phase factor for electrons or photons (if their energies are the same), and the solution to Schrodinger's equation is exactly the same in either case, as was explicitly verified.

The preceding discussion shows that the dynamic behavior of the system is exactly the same for photons in the presence as a strong Zeno effect as it is for non-interacting fermions (with no Zeno effect). Thus a coupled-wave guide device of this kind would also implement the $\sqrt{SWAP'}$ operation if two non-interacting electrons were used, including the factor of $i$ phase shift.

If the photons truly behave like fermions, then we might expect their creation and annihilation operations to obey anti-commutation relations. In order to investigate this situation, we apply a unitary transformation to the interaction picture in which the Zeno effect is included in the dressed-state operators. As usual, the total Hamiltonian can been expressed as the sum of two parts:

$$H = H_0 + H_{\text{int}} \tag{23}$$

Here we choose $H_0$ to include the unperturbed photon energies as well as the two-photon absorption, while $H_{\text{int}}$ includes only the coupling between the two wave guides as given in Eq. (2). An operator $\hat{O}_S$ in the Schrodinger picture is then transformed as usual into a time-dependent operator $\hat{O}(t)$ in the interaction picture [35] given by

$$\hat{O}(t) \equiv e^{iH_0t/\hbar} \hat{O}_S e^{-iH_0t/\hbar}. \tag{24}$$

We will initially consider a single mode in a single fiber and, in order to avoid confusion, we will denote the dressed operators $\hat{a}(t)$ and $\hat{a}^\dagger(t)$ by $\hat{A}(t)$ and $\hat{A}^\dagger(t)$, while the Schrodinger-picture operators will be written simply as $\hat{a}$ and $\hat{a}^\dagger$.

We first evaluate the equal-time commutation properties of the dressed operators:

$$\begin{aligned}
[\hat{A}(t), \hat{A}^\dagger(t)] &= \hat{A}(t)\hat{A}^\dagger(t) - \hat{A}^\dagger(t)\hat{A}(t) = \\
&(e^{iH_0t/\hbar}\hat{a}e^{-iH_0t/\hbar})(e^{iH_0t/\hbar}\hat{a}^\dagger e^{-iH_0t/\hbar}) - (e^{iH_0t/\hbar}\hat{a}^\dagger e^{-iH_0t/\hbar})(e^{iH_0t/\hbar}\hat{a}e^{-iH_0t/\hbar}) = \\
&e^{iH_0t/\hbar}[\hat{a}, \hat{a}^\dagger]e^{-iH_0t/\hbar} = [\hat{a}, \hat{a}^\dagger]
\end{aligned} \tag{25}$$

Here we have used the fact that the commutator is a non-operator (a number) that commutes with the time evolution operator. Thus we see that the Zeno effect cannot change the equal-time commutation properties of the field operators.

At first these results may seem puzzling, since they seem to imply that the photons obey commutation relations even though their dynamic evolution is that of a fermion. The results of Eq. (25) are due to the fact that $\hat{A}^\dagger(t)$ instantaneously creates a photon, after which $\hat{A}(t)$ immediately annihilates it, whereas the Zeno effect requires a finite time interval in order to take effect. But the dynamics of the system really depend on time-averaged operators, as can be seen from perturbation theory [35]:

$$|\psi(t)\rangle = |\psi(0)\rangle + \frac{1}{i\hbar}\int\limits_0^t dt'\, H_{\text{int}}(t')|\psi(0)\rangle + \frac{1}{(i\hbar)^2}\int\limits_0^t dt'\int\limits_0^{t'} dt''\, H_{\text{int}}(t')H_{\text{int}}(t'')|\psi(0)\rangle + \dots \quad (26)$$

This motivates us to consider the time-averaged product of two operators $\hat{A}(t)$ and $\hat{B}(t)$ over the time interval $\tau$, which we define as

$$\overline{\hat{A}\hat{B}} \equiv \frac{2}{\tau^2}\int\limits_0^\tau dt'\int\limits_0^{t'} dt''\, \hat{A}(t')\hat{B}(t'') \quad (27)$$

For simplicity, we will assume that the photons are emitted or absorbed on resonance with an atomic transition, which is equivalent to neglecting the photon energies.

In order to evaluate the time-averaged commutation properties, we note that the first-order probability amplitude $c_1$ to emit a second photon, given that one already exists, vanishes in the limit of a strong Zeno effect:

$$c_1 = \frac{1}{i\hbar}\int\limits_0^\tau dt'\, \hat{A}^\dagger(t')|1\rangle \rightarrow 0 \quad (28)$$

Here $|1\rangle$ denotes a state with a single photon in the optical fiber of interest. It follows from this and Eq. (27) that

$$\overline{\hat{A}\hat{A}^\dagger}|1\rangle = 0 \quad (29)$$

in the limit of a strong Zeno effect. In a similar way, it can also be shown that

$$\overline{\hat{A}\hat{A}^\dagger}|0\rangle = 1$$
$$\overline{\hat{A}^\dagger\hat{A}}|1\rangle = 1 \quad (30)$$
$$\overline{\hat{A}^\dagger\hat{A}}|0\rangle = 0$$

It immediately follows from Eqs. (29) and (30) that

$$\overline{\{\hat{A},\hat{A}^\dagger\}} = 1 \quad (31)$$

when acting on the subspace of states ($|0\rangle$ and $|1\rangle$) that are allowed by the Zeno effect. Thus we conclude that the time-averaged creation and annihilation operators obey anti-commutation relations within the allowed subspace, which is consistent with the other properties of the system.

Up to this point, we have considered the commutation properties of the operators in a single optical fiber. If, instead, if we consider the field operators for two different fibers, $i$ and $j$, then it is apparent that

$$\overline{[\hat{A}_i, \hat{A}_j^\dagger]} = 0 \qquad (i \neq j) \tag{32}$$

This result is consistent with an interpretation in which the dressed photons in different optical fibers behave as if they were different kinds of fermions (as would be the case for protons vice electrons, for example). As a result, interchanging two optical fibers containing two photons under the influence of the Zeno effect as in Fig. 6a does not introduce a minus sign such as the one in Eq. (11). The dressed-state photons can still be viewed as fermions, however, but with a different species in each fiber.

To summarize the results of this section, we have shown that photons in the presence of a strong Zeno effect of this kind are subject to the Pauli exclusion principle, that the HOM interference effect corresponds to that of fermions, that the dynamics of the system are identical to those of non-interacting fermions, and that the time-averaged field operators obey anti-commutation relations. In that sense we can say that the photons behave like non-interacting fermions.

## VI.    A new paradigm for quantum computing

Our conclusion that the photons behave like non-interacting fermions in Zeno gates may seem contradictory at first, since there are no-go theorems that show that quantum computation cannot be performed using only non-interacting fermions [18, 19]. Here we show that our results are consistent with the various no-go theorems for both fermions and bosons [36, 37]. In the process, we describe a new paradigm for quantum computation.

We first consider what would happen if non-interacting fermions, such as electrons, were incident on the controlled-Z (controlled phase) gate of Fig. 6b. Once again, we assume that there is no Zeno effect and the fermions are truly non-interacting. As we showed in Section V, the coupled wave guide device of Fig. 3b will implement the *SWAP'* operation just as it does for photons in the presence of the Zeno effect. However, interchanging the paths of two electrons as in Fig. 6a will implement the *SWAP'* operation instead of *SWAP* due to the anti-commutation relations obeyed by the electrons. This can be seen in more detail by assuming that the system is initially in a state given by

$$|\psi_0\rangle = \hat{b}_i^\dagger \hat{b}_j^\dagger |0\rangle \tag{33}$$

which corresponds to one particle in each of the two wave guides. Interchanging the two paths will then produce a state $|\psi'\rangle$ that is given by

$$|\psi'\rangle = \hat{b}_j^\dagger \hat{b}_i^\dagger |0\rangle = -|\psi_0\rangle \qquad (34)$$

The minus sign in the above equation shows that the *SWAP'* operation will be implemented when two wave guides containing two identical fermions are interchanged. Thus the device shown in Fig. 6b will produce two *SWAP'* operations, which implements the identity operator instead of a controlled phase gate. This shows that our calculations are consistent with the no-go theorems for non-interacting fermions [18, 19].

There are also no-go theorems for non-interacting bosons [36, 37]. The ability to perform quantum computation using single photons and the Zeno effect is due to the fact that the photons can behave like non-interacting fermions in one part of a circuit and like non-interacting bosons in other parts of the circuit, as illustrated in Fig. 9. Here the particles are forced to behave like fermions while their paths are interchanged, which implements the *SWAP'* operation. The particles are then forced to behave like bosons while the paths are interchanged again, which implements the *SWAP* operation. The combination of the two operations implements the controlled phase gate as in Fig. 6b, which is universal for quantum computation. The operations illustrated in Fig. 9 can be viewed as a new paradigm for quantum computation

It has been suggested that Eq. (34) (and the conclusions of this paper) must be incorrect because two electrons located at different positions are "distinguishable", in which case "statistics cannot play a role". In fact, any two electrons are identical regardless of their locations, and it is a fundamental postulate of quantum field theory [38] that a fermionic field operator $\hat{\psi}^\dagger(\vec{r},t)$ must satisfy the anti-commutation relations

$$\{\hat{\psi}^\dagger(\vec{r},t), \hat{\psi}^\dagger(\vec{r}',t)\} = 0 \qquad (35)$$

Here the operators $\hat{\psi}^\dagger(\vec{r},t)$ and $\hat{\psi}^\dagger(\vec{r}',t)$ create identical fermions at two different locations $\vec{r}$ and $\vec{r}'$, as in the argument stated above. Thus the minus sign in Eq. (34) is correct, however counter-intuitive it may seem, and it becomes observable if the two paths are brought back together to produce interference effects. This type of confusion can be avoided by thinking in terms of "identical" particles rather than "indistinguishable" particles.

## VII.    Two-qubit encoding and scalability

A scalable approach to quantum computing would require the probability $P_E$ of an error in a logic operation to be below the threshold for quantum error correction. The results of Section III show that $P_E$ only decreases as $\pi^2/4N$, where $N$ is the number of measurements used to implement the Zeno effect. This would seem to imply that a large number of measurements or a large two-photon absorption rate would be required for scalable operation. Fortunately, the two-qubit encoding [1] of Knill, Laflamme, and Milburn (KLM) can be used to greatly reduce the value of $P_E$ for a relatively small value of $N$.

In the original KLM approach [1] to linear optics quantum computing, the failure events correspond to situations in which the value of a qubit is measured in the computational basis. They showed that measurement failures of this kind could be corrected if the logical qubits $|0_L\rangle$ and $|1_L\rangle$ were encoded in the quantum states of two photons in the following way:

$$\begin{aligned} |0_L\rangle &\equiv |00\rangle + |11\rangle \\ |1_L\rangle &\equiv |01\rangle + |10\rangle \end{aligned}$$

(36)

Here $|00\rangle$ denotes a state with two physical qubits (photons), $q_1$ and $q_2$, with logical values of 0, and similarly for the remaining states in Eq. (36). If the value of one of the physical qubits, say $q_1$, is measured, then the logical qubit can be restored to its original value by replacing $q_1$ with an appropriate superposition state and then applying a CNOT operation between the two qubits [1].

In our Zeno gate approach, failure events can occur if the strength of the two-photon absorption is not sufficiently large. In that case, the emission of two photons into the same fiber core will not be totally inhibited and two photons may actually be absorbed with a small probability. This process will measure the values of both qubits to be 1 provided that such events can be observed when they occur. For example, a two-photon absorption event could be observed if one or both of the secondary photons were detected in the four-wave mixing process of Fig. 8c. This differs from the original KLM approach in that both qubits are measured at the same time, which gives a somewhat lower error threshold as we shall see.

In this two-qubit encoding, a CNOT operation between two logical qubits $q$ (the control) and $q'$ (the target) can be applied by applying two CNOT operations to the physical qubits, as illustrated in Fig. 10. One CNOT operation is applied between $q_1$ of the control and $q_1'$ of the target, while the second CNOT operation is applied between $q_2$ of the control and $q_1'$ of the target. We will assume that an individual CNOT operation between two of the physical qubits fails with a probability $P_F$, and we will estimate the corresponding probability $P_F'$ that the operation will fail at the upper, logical level. Scalability requires that $P_F' < P_F$, in which case the encoding could be concatenated [24] to produce an arbitrarily small error in a CNOT operation at the logical level. To simplify the analysis, we assume that $P_F << 1$.

In order for a failure to occur at the upper logical level, there must first be a failure in one of the lower-level CNOTs. If the first *CNOT* operation fails with probability $P_F$, that will measure qubits $q_1$ and $q_1{}'$, but they can be corrected using a subsequent CNOT operation for each. The probability that one of these subsequent CNOT operations will fail is $2P_F$, so that the overall probability of a failure at the logical level due to this mechanism is $2P_F{}^2$.

An overall error can also occur if the second *CNOT* operation fails, which will measure qubits $q_2$ and $q_1{}'$. The probability of a failure at the upper logical level after an attempt at correcting the qubits is once again $2P_F{}^2$, so that the total failure probability at the upper logical level is $4P_F{}^2$. Scalability requires that $P_F{}' < P_F$, or $4P_F{}^2 < P_F$, which shows that the threshold for the correction of these kinds of errors is $P_F < 1/4$.

This two-qubit encoding cannot correct for more general kinds of errors, such as bit flips or erasure errors. As a result, the two-qubit encoding would have to be incorporated into the lowest level of a more general encoding [24]. In addition, the total probability for an error other than a measurement error within the two-qubit encoding must be smaller than the error threshold of the higher-level encoding. If the two-photon absorption is sufficiently strong, then it may be possible to use only two or three layers of the two-qubit encoding while minimizing other sources of errors. This is easier to achieve using Zeno gates than it is in the original KLM approach because the latter requires a large number of encoding layers along with the successful preparation and detection of a large number of ancilla photons.

## VIII.   Summary and conclusions

We have shown that the quantum Zeno effect can be used to suppress the failure events that would otherwise occur in a linear optics approach to quantum computing. The use of Zeno gates of this kind would avoid the need for large numbers of ancilla photons and high-efficiency detectors, and the logic devices would become deterministic in the limit of a strong Zeno effect. We have also shown that a two-qubit encoding can be used to achieve small error rates even when the Zeno effect has limited strength, which is an important consideration for a scalable approach to quantum computing.

As is usually the case in the Zeno effect, no actual measurements or observations are required and the same results can be obtained using strong two-photon absorption to inhibit the emission of more than one photon into the same optical fiber. We have estimated the achievable rate of two-photon absorption in optical fibers and we are optimistic that single-photon scattering and absorption can be reduced to a sufficiently small level for these devices to be useful in practical applications. The feasibility of the approach can only be demonstrated by performing the relevant experiments, one of which is now in progress.

The Zeno effect has previously been proposed as a method of quantum error correction [39-41] or error reduction [42-43], and it is similar to the "bang-bang" method of error reduction [44-47].   As we mentioned above, a somewhat similar use of decoherence to suppress unwanted error states in cavity QED logic gates has been discussed by Beige et al. [25-27].   Although these earlier approaches obviously have much in common with our proposal for Zeno gates, they all make use of physical interactions to perform the basic logic operations, combined with the Zeno effect to reduce the residual error rate.   In our approach, there are no physical interactions between the photons and the Zeno effect is totally responsible for the ability to perform the logic operations.

From a basic physics perspective, we have shown that the photons behave like non-interacting fermions instead of bosons in the presence of a strong Zeno effect.   The photons are forced to obey the Pauli exclusion principle in the limit of a strong Zeno effect and their interference properties in a Hong-Ou-Mandel interferometer are characteristic of fermions instead of bosons.   The dynamic evolution of the system is the same as it would be for non-interacting fermions, and the time-averaged creation and annihilation operators obey anti-commutation relations instead of commutators.   The operation of Zeno logic gates depends on the fact that the photons can be forced to behave like fermions in one part of a circuit and then forced to behave like bosons in other parts of the circuit, which gives rise to a new paradigm for quantum computing as illustrated in Fig. 9.

In summary, we conclude that the use of the quantum Zeno effect to suppress the failure events in linear optics quantum computing may be of practical use in quantum information processing and it exhibits a fundamentally new type of behavior in which photons behave as if they were fermions.

We would like to acknowledge valuable discussions with C.W.J. Beenaker, Michael Fleischhauer, Mark Heiligman, Gershon Kurizki, Gerard Milburn, Tim Ralph, Ian Walmsley, and Colin Williams.  This work was supported by ARDA, ARO, ONR, NSA, and IR&D funding.

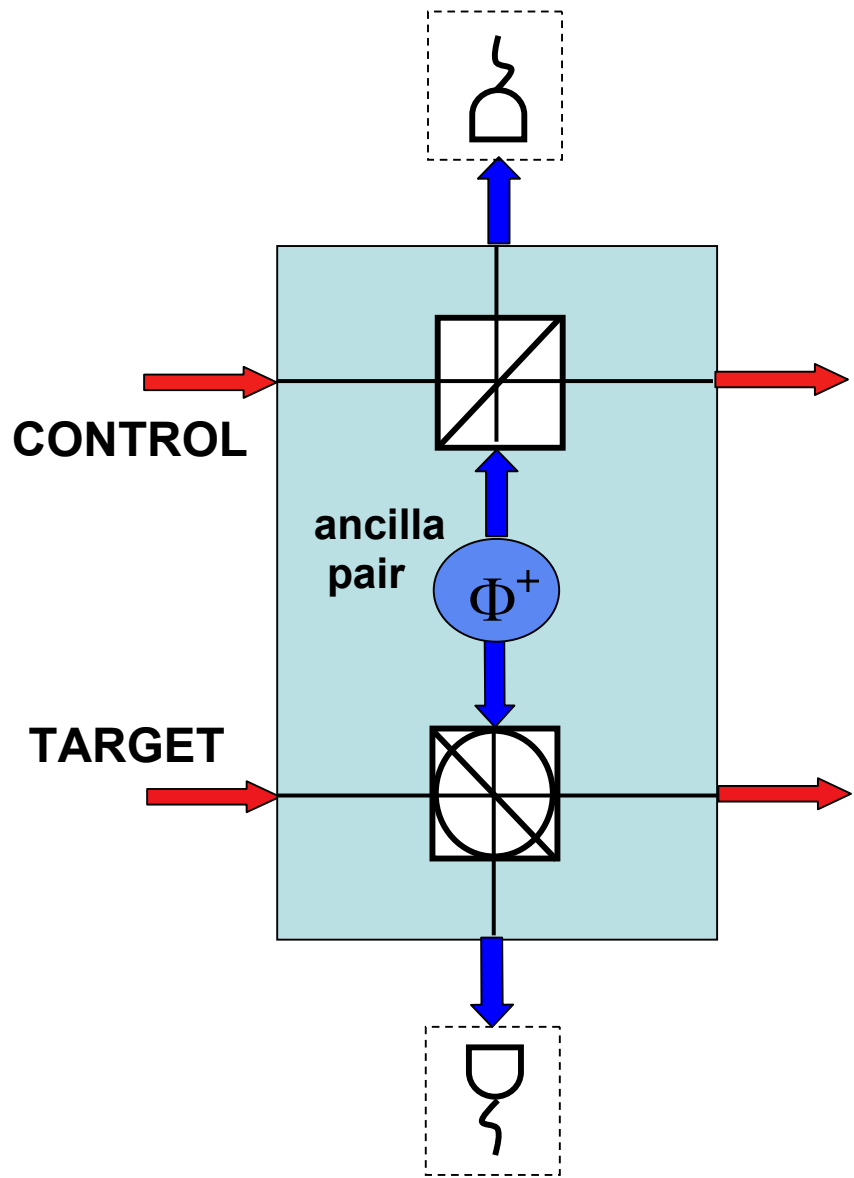

Fig. 1. Implementation of a controlled-NOT logic gate that succeeds with probability ¼. All of the failure events correspond to situations in which two photons were emitted into the same optical fiber. A more detailed description can be found in Ref. [22].

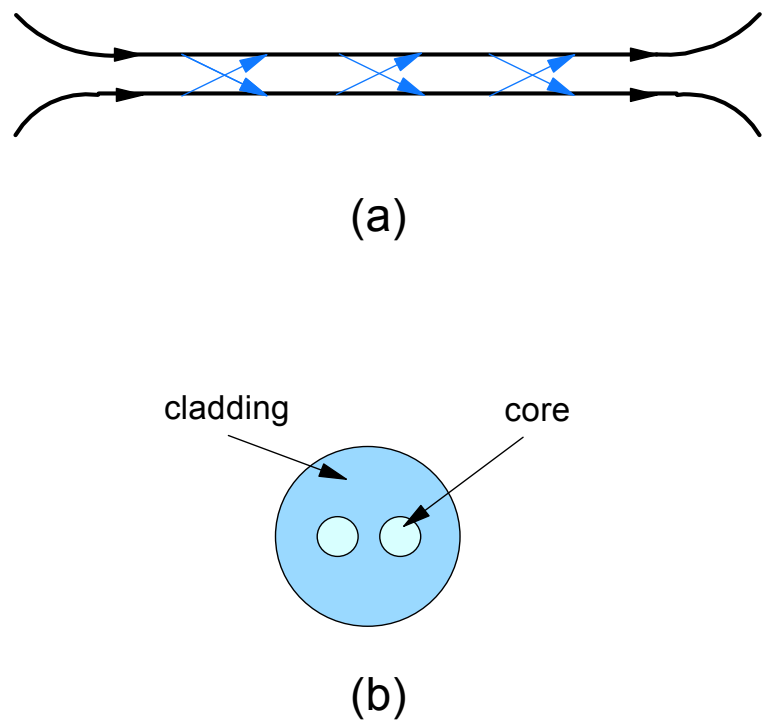

Fig. 2. Coupled optical fibers used to implement the beam splitting process in a continuous way, which allows the use of the Zeno effect. (a) Side view showing the coupling between two optical fiber cores via their evanescent field, which is equivalent to the tunneling of photons from one core to the other. (b) End view showing a dual-core optical fiber that could be used to implement such a beam splitter. Devices of this kind are commercially available.

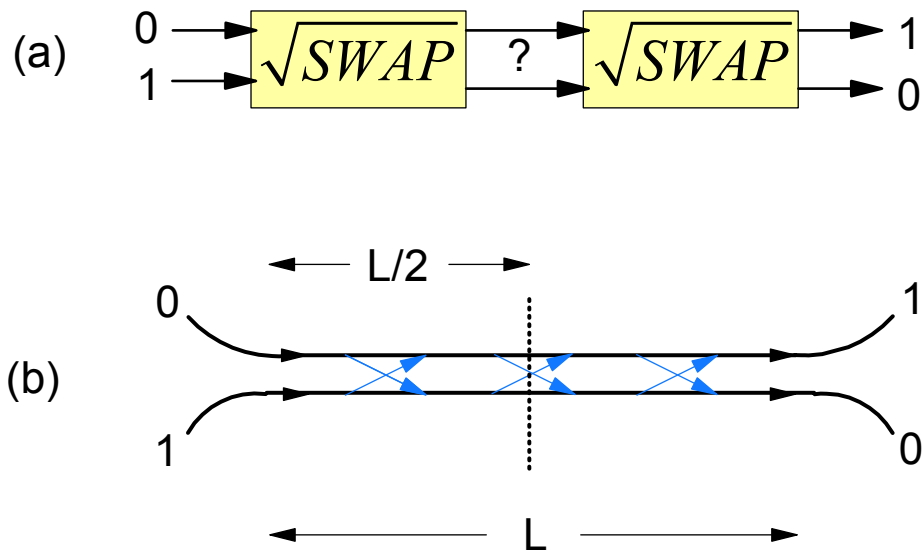

Fig. 3. (a) Operation of the $\sqrt{SWAP}$ gate, which is defined in such a way that it swaps the values of any two logical inputs when it is applied twice (squared). If the operation is only applied once, however, a quantum superposition of states is created and neither qubit has a well-defined value, as illustrated by the question mark. (b) Potential implementation of a $\sqrt{SWAP}$ gate by bringing the cores of two optical fibers in close proximity, which allows a photon in one core to be coupled into the other core. If length L produces a SWAP operation, then half that length (dashed line) will produce the $\sqrt{SWAP}$ operation, aside from errors that can be suppressed using the Zeno effect and a phase factor discussed in the text.

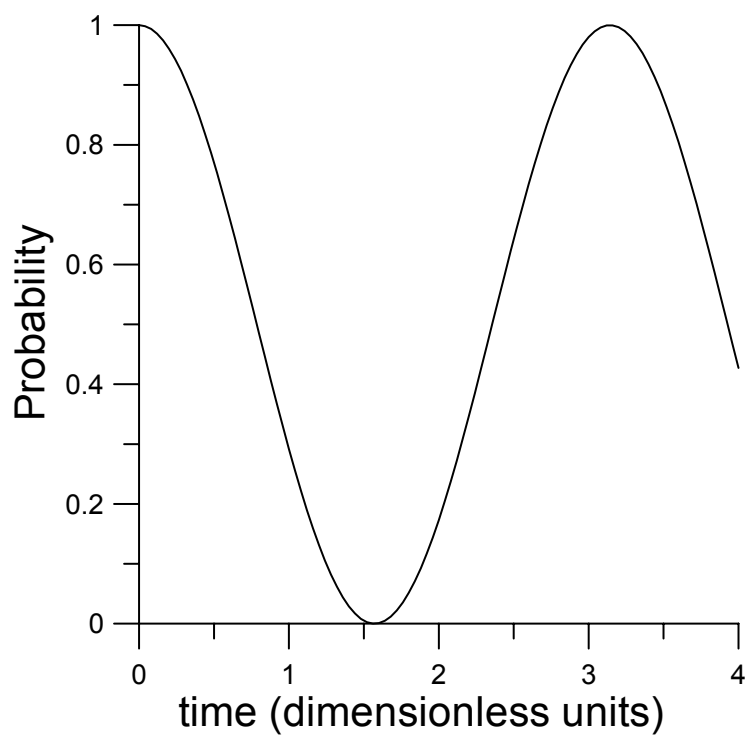

Fig. 4. Probability that a single photon will be found in path 1 of the coupled optical fiber device of Figs. 2 and 3. A single photon was assumed to be incident in path 1 and Schrodinger's equation was solved using the Hamiltonian of Eq. (2). These results are equivalent to Rabi oscillations in a two-level atom.

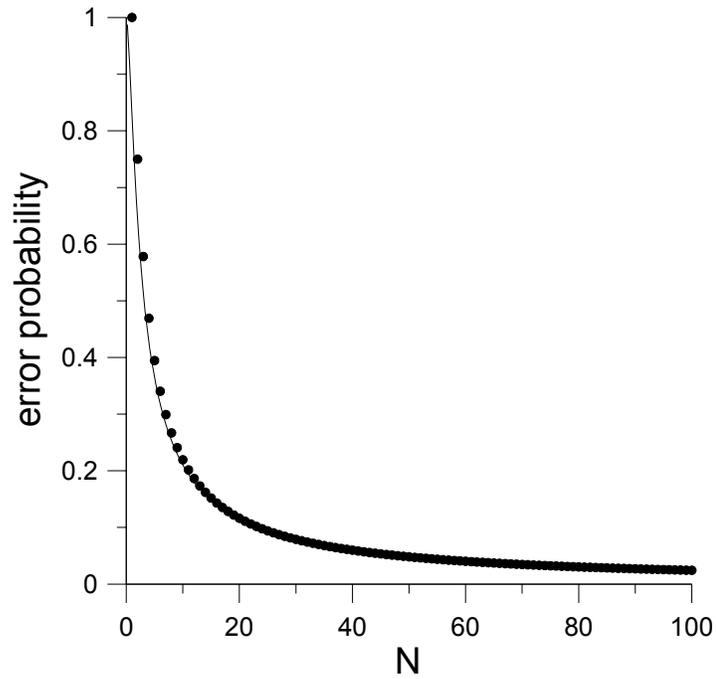

Figure 5. Probability $P_E$ of an error in the output of the $\sqrt{SWAP}'$ gate when a photon is incident in both input modes. The dots correspond to the value of $P_E$ as a function of the number $N$ of equally-spaced measurements made to determine the presence of two photons in the same optical fiber core. The solid line corresponds to similar results obtained in the presence of two-photon absorption, where the parameter $N$ is then defined as $N = \Delta t / 4\tau_D$, with $\Delta t$ the interaction time and $\tau_D$ the two-photon decay time.

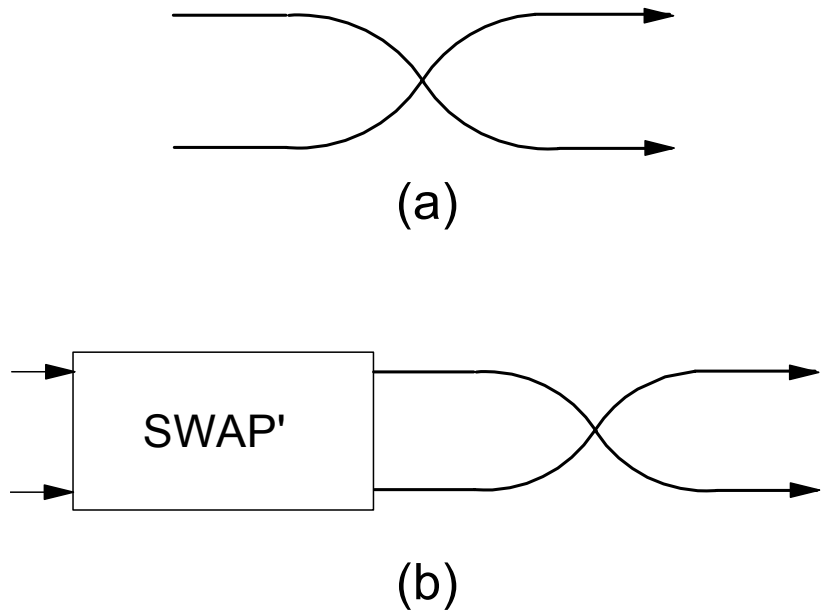

(a)

(b)

Fig. 6.  (a) Implementation of a *SWAP* operation for photons by simply crossing two optical fibers.  (b) A controlled-Z gate for photons constructed by applying the *SWAP*′ operation of Eq. (11) followed by the *SWAP* operation illustrated above.  This circuit produces only the identity operator for non-interacting fermions.

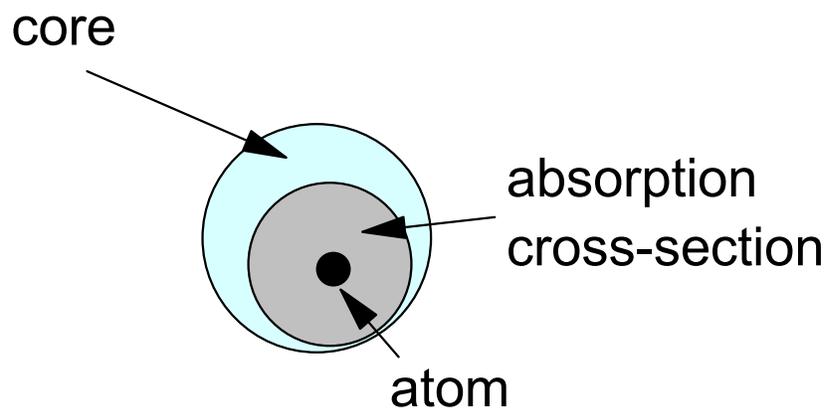

Fig. 7. Expanded view of one of the optical fiber cores containing a single atom. The cross-section for the absorption of a resonant photon can be comparable to the area of the core itself, in which case there can be a large interaction between two photons and a single atom. The effects of detuning and collisions must also be taken into account, but those effects can be compensated to give a two-photon absorption rate comparable to that from this simple picture.

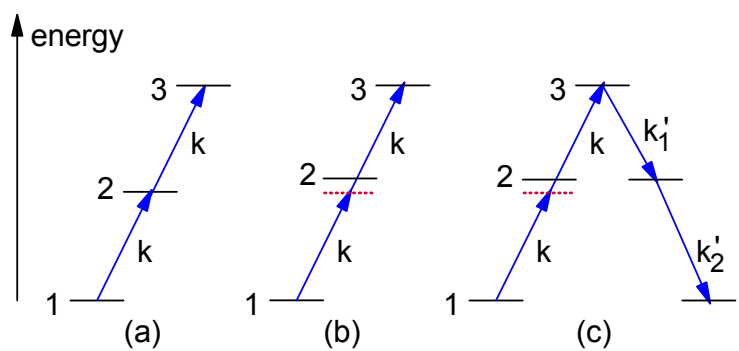

Figure 8. Energy-level diagrams illustrating the use of a three-level atom to produce two-photon absorption. The atomic levels are labeled 1 through 3 while the photons are represented by the arrows. (a) Successive absorption of two photons whose energies are on resonance with that of the atomic transitions. (b) Elimination of single-photon absorption by increasing the energy of level 2, in which case only two-photon absorption satisfies energy conservation. (c) A virtual process in which the two original photons with wave vector $k$ are absorbed, followed by the re-emission of two different photons with wave vectors $k_1{}'$ and $k_2{}'$.

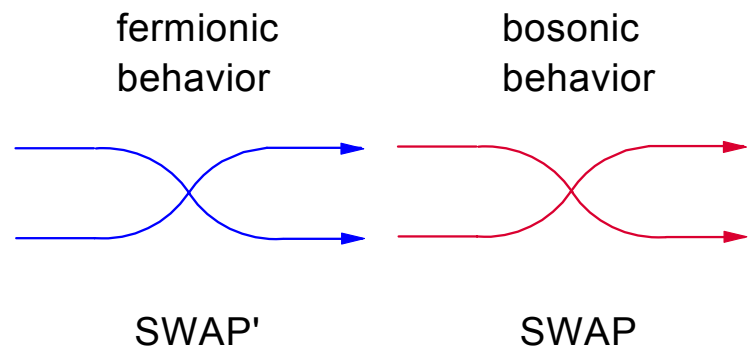

Fig. 9. A new paradigm for quantum computation. It is assumed that a set of particles can be forced to behave like fermions while their paths are interchanged, which implements the *SWAP*' operation of Eq. (11). The particles are then forced to behave like bosons while the paths are interchanged again, which implements the *SWAP* operation. The combination of these two operations implements a controlled-Z gate (controlled phase gate), which is universal for quantum computation.

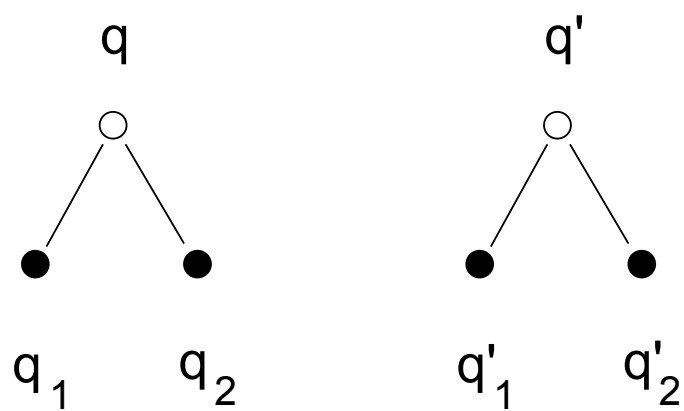

Fig. 10. A two-qubit encoding that can be used to correct for failure events in which a Zeno effect of limited strength allows the absorption of two photons. Logical qubit $q$ is encoded in the values of two physical qubits (photons) $q_1$ and $q_2$, and similarly for logical qubit $q'$.